\documentclass[manuscript]{acmart}
\usepackage{graphicx}
\usepackage{subfig}
\usepackage{tabularx}
\usepackage [autostyle, english = american]{csquotes}
\MakeOuterQuote{"}
\AtBeginDocument{%
  \providecommand\BibTeX{{%
    \normalfont B\kern-0.5em{\scshape i\kern-0.25em b}\kern-0.8em\TeX}}}

\acmConference[AutomotiveUI '21 Adjunct]{13th International Conference on Automotive User Interfaces and Interactive Vehicular Applications}{September 9--14, 2021}{Leeds, United Kingdom}
\acmBooktitle{13th International Conference on Automotive User Interfaces and Interactive Vehicular Applications (AutomotiveUI '21 Adjunct), September 9--14, 2021, Leeds, United Kingdom}
\acmDOI{10.1145/3473682.3480282}
\acmISBN{978-1-4503-8641-8/21/09}

\begin{document}

\title[An Autonomous Driving System]{An Autonomous Driving System - Dedicated Vehicle for People with ASD and their Caregivers}

%\author{Anonymous Author(s)}
\author{Gandhimathi Padmanaban}
\email{gmathi@umich.edu}
\author{Nathaniel Jachim}
\email{njachim@umich.edu}
\author{Hala Shandi}
\email{hshandi@umich.edu}
\author{Lilit Avetisyan}
\email{lilita@umich.edu}
\author{Garrett Smith}
\email{gssm@umich.edu}
\author{Howraa Hammoud}
\email{howraah@umich.edu}

\author{Feng Zhou}
\email{fezhou@umich.edu}
\orcid{0000-0001-6123-073X}
\affiliation{  \institution{University of Michigan, Dearborn} \streetaddress{4901 Evergreen Road} \city{Dearborn} \state{MI} \country{USA} \postcode{48128}}

%\renewcommand{\shortauthors}{Anonymous, et al.}
\renewcommand{\shortauthors}{Padmanaban, et al.}

\begin{abstract}
{Automated driving system - dedicated vehicles (ADS-DVs), specially designed for people with various disabilities, can be beneficial to improve their mobility. However, research related to autonomous vehicles (AVs) for people with cognitive disabilities, especially Autism Spectrum Disorder (ASD) is limited. Thus, in this study, we focused on the challenge that we framed: “How might we design an ADS-DV that benefits people with ASD and their caregivers?”. In order to address the design challenge, we followed the human-centered design process. First, we conducted user research with caregivers of people with ASD. Second, we identified their user needs, including safety, monitoring and updates, individual preferences, comfort, trust, and reliability. Third, we generated a large number of ideas with brainstorming and affinity diagrams, based on which we proposed an ADS-DV prototype with a mobile application and an interior design. Fourth, we tested both the low-fidelity and high-fidelity prototypes to fix the possible issues.  Our preliminary results showed that such an ASD-DV would potentially improve the mobility of those with ASD without worries.}
\end{abstract}



\keywords{Autonomous vehicle, autism spectrum disorder, human-centered design, dedicated vehicles.}

\maketitle

\section{Introduction}
According to the U.S. Center for Disease Control (CDC), one in 59 (or 1.7\%) of children have autism, and almost half of those with autism have average to high levels of intelligence \cite{baio2018prevalence}. However, it is difficult for them to independently transport from one place to another, which is not only important to access jobs, healthcare, and other critical destinations, but also allows for participation in vocational and leisure purposes \cite{christiansen2005occupational}. A few studies \cite{classen2013driving,almberg2017experiences} have highlighted that upon reaching the legal age, individuals with autism spectrum disorder (ASD) are indeed interested in driving and that many of them are either learning to drive or are current drivers. However, only 34 \% of people with ASD received a driving license \cite{curry2018longitudinal}, which indicates that two thirds of people with ASD need are dependent on someone for their ride.

People with ASD have problems with anxiety, multitasking, motor coordination, attention allocation, tolerating unexpected changes in driving routes and understanding non-verbal communication of other road-users which makes self-driving and public transportation a big hurdle for them \cite{michon1985critical,classen2013driving}. Driving is a multidimensional activity which requires competencies in the operational, tactical and strategic levels of driving \cite{michon1985critical}. Thus, driving becomes more vulnerable for people with ASD, especially when it involves social activities, such as non-verbal interactions, through hand gesturing and eye contact, with other road users to indicate intentions. Previous studies reveal that many of the diagnostic factors associated with autism may contribute to driving difficulties, such as problems in executive functions, social-cognitive, motor, sensory perception, and integration of sensory-motor skills \cite{almberg2017experiences,cox2012brief}. For example, Sheppard et al. \cite{sheppard2010brief} reported that people with ASD identified fewer social hazards (e.g., pedestrians, cyclists) and slower reaction times to all hazards compared to the control group.

Currently, conventional vehicles do not provide any adaptable options to meet the needs of individuals with ASD and their caregivers. Depending on the behavior of people with ASD in overwhelmed moments, caregivers have to get off the driving task, causing risky situations for other road users, and concentrate on them in order to prevent them from self-harming. Automated driving system - dedicated vehicles (ADS-DVs), especially those designed and operated at SAE Levels 4-5 automated driving systems (ADS) (geo-fenced and fully AVs) \cite{sae2016definitions}, can improve their mobility by removing the driving task \cite{ayoub2019manual,ayoub2020otto} and thus produce positive benefits for people with ASD. We did not consider SAE Level 3 AVs due to the fact that these vehicles would still require the driver to take over control from automated driving to manual driving when the system reached the functional limit (see \cite{du2020psychophysiological,zhou2020driver,du2020predicting,zhou2021using}). Such takeover transitions are especially challenging due to the possible issues in sensory-motor skills of people with ASD \cite{almberg2017experiences,cox2012brief}.

Furthermore, the U.S. Department of Transportation also promotes innovative design solutions in the ADS-DV field which could enable people with physical, sensory, and cognitive disabilities to use vehicles, especially for SAE Levels 4-5 AVs. This need has been further highlighted by the current COVID-19 pandemic \cite {zhou2020not,ayoub2021combat} for vulnerable populations, such as those with ASD, to have on-demand transportation services to access healthcare, pharmacies, grocery stores, and other essential services. Innovations in ADS-DV differ from technological challenges to non-standard interiors and adapted equipment since regular vehicles cause challenges in these stacks as well, where no special needs are considered in vehicle design. Therefore, it is of great importance to design ADS-DV for people with ASD in order to improve their mobility.

In order to understand how ADS-DVs at SAE Levels 4-5 can be beneficial for people with ASD and what the caregivers’ requirements of the ADS-DV are in order to trust and allow a person with ASD to ride in it alone, the objectives of this study are 1) to explore behavioral patterns, needs, and comfort issues of those with ASD in a regular vehicle, 2) to understand caregivers' concerns about people with ASD during the ride, and 3) to develop a new ADS-DV prototype, which can help solve the most pressing barriers for people with ASD during the ride to improve their mobility. 

\section{Methods and Results}
In order to achieve the research objectives, we adopted Stanford d.school’s design thinking model \cite{reference_9} and went through the following design processes: 1) empathize - to understand mobility issues about people with ASD deeply, 2) define - to analyse the collected information and define the problems, 3) ideate - to propose a solution for the identified problem, 4) prototype - to bring the solution to the material level, and 5) test - to evaluate the prototype in terms of user satisfaction and effectiveness for people with ASD. 
%In order to achieve the research objectives, we used a human-centered design process adopted from the Stanford d.school, including empathize, define, ideate, prototype, and test \cite{reference_9}, as described below. 

\subsection{Empathize}
To understand the behavioral patterns, needs, and issues of people with ASD during a ride, we interviewed three participants through Zoom (San Jose, California, U.S., https://zoom.us/), including two behavioral technicians who were responsible for the care of people with ASD during transportation and one parent of a child with ASD. During the interviews, we asked our participants to talk about their experience by sharing their daily drives and struggles involved in transporting people with ASD. They also answered questions related to the environment and the design of the vehicle. By these interviews, it was confirmed that people with ASD did have mobility issues and it was often not enough to drive them from one place to another with a single driver (e.g., parent or a behavioral technician) due to problems, feelings, and struggles during the transportation. Example quotes included ``\emph{He loves the car, keeps moving, [but] he hates it when we have to stop when it is slow or busy. Sometimes I speed because he requests that, but then for safety concerns I slow down and he gets mad again}" (i.e., problems), ``\emph{Seats and belts in the car and the safety thing. He does not like to be sitting in the chair and he does not like the set belt. He get mad when I force him to use the belt and he start crying, screaming and hitting his head with the windows to get attention, so my dream was always to have a safer windows because i am afraid that he breaks the glasses one day}" (i.e., feelings and struggles).

To get further information, we decided to enlarge our sample size by creating a survey through Qualtrics (Provo, Utah, U.S., https://www.qualtrics.com/) with a total of 23 questions that helped us to understand the nature of the design challenge and their attitudes towards AVs. We used social media (Facebook, LinkedIn, and Instagram) to recruit a total of 31 participants (6.5\% with ASD, 38.7\% family members, 41.9\% friends, and 12.9\% others) who had experience in taking care of or sharing a ride with people with ASD. The survey mainly explored the questions: 1) how do those with ASD behave differently during the rides compared to those without ASD, 2) how often do such rides occur in a week, 3) what are the possible consequences, 4) how much do they trust in AVs, and 5) what are features they would like to see in such an ADS-DV for transporting those with ASD. The summary of the results the survey is shown in Table 1.

\begin{table*} [bt!]
\caption{Summary of the results from the survey}
\label{tab:issue}
\centering
\begin{tabularx}{\textwidth} {lX}
    \hline
    Items & Summary of the results\\
    \hline
    Different behaviors from others & Overreaction (e.g., feeling excited, anxious, showing maximum behavior like tantrums) to vehicle behavior and scenery (e.g., driving speed, honks, lights outside, familiar restaurants)\\
    Frequency of rides & M = 3.10, SD = 1.72 per week\\
    Possible consequences & Too excited or anxious; self-injuring behaviors; messing with the vehicle; removing seat belts, even attempting to open the door \\
    Trust in AVs & M = 2.81, SD = 1.30 (5-point Likert scale, where 1 = definitely not trust and 5 = definitely trust)  \\
    Preferred features for ADS-DV & comfortable chairs; features to keep them entertained; communication with families; specialized interface; constraints to protect them (from self harm)\\
    \hline 
    
\end{tabularx}
\end{table*}

\subsection{Define}
\begin{figure}[bt!]
  \centering
  \includegraphics[width=0.9\linewidth]{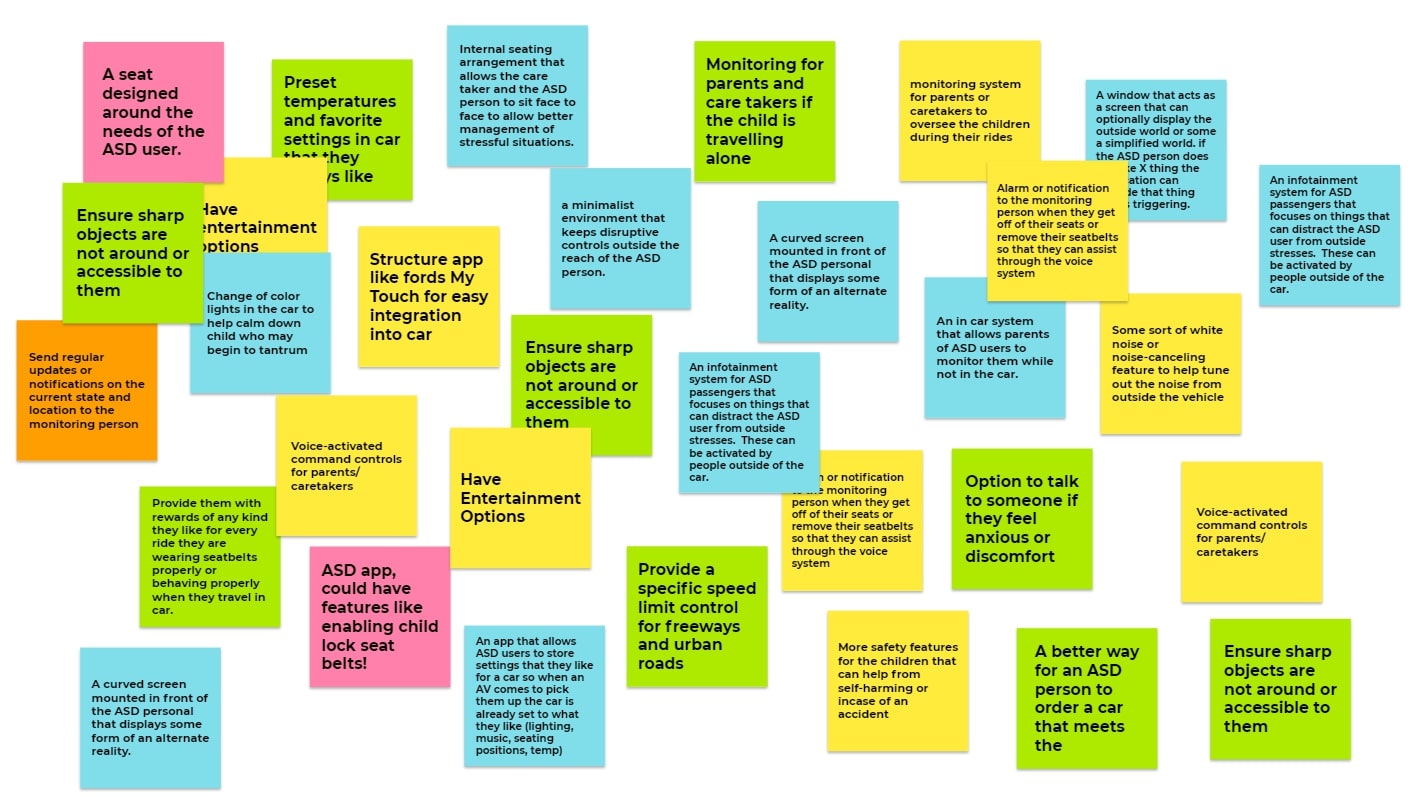}
  \caption{Affinity Diagram using Google Jamboard.}
  \label{fig:Affinity}
  \vspace{-10pt}
\end{figure}

In the define stage, we made use of affinity diagram in Google Jamboard (California, U.S., https://jamboard.google.com) to create different categories of the data collected from the survey and three interviews (see Figure \ref{fig:Affinity}). First, we identified two target user groups for our design challenge, i.e., caregivers of people with ASD and those with ASD. Within each target user group, we were able to further break down the data collected into different themes and patterns. Then, from the caregiver’s perspective, we found that a safe ride that would protect those they were caring was the most important, followed by monitoring and updates of the passenger with ASD and comfort of, trust in, and reliability of the AV. From the perspective of those with ASD, we found that safety, mobility, anxiety, individual preferences were among the most important categories. 

\begin{figure*}[bt!]
  \centering
  \includegraphics[width=0.9\linewidth]{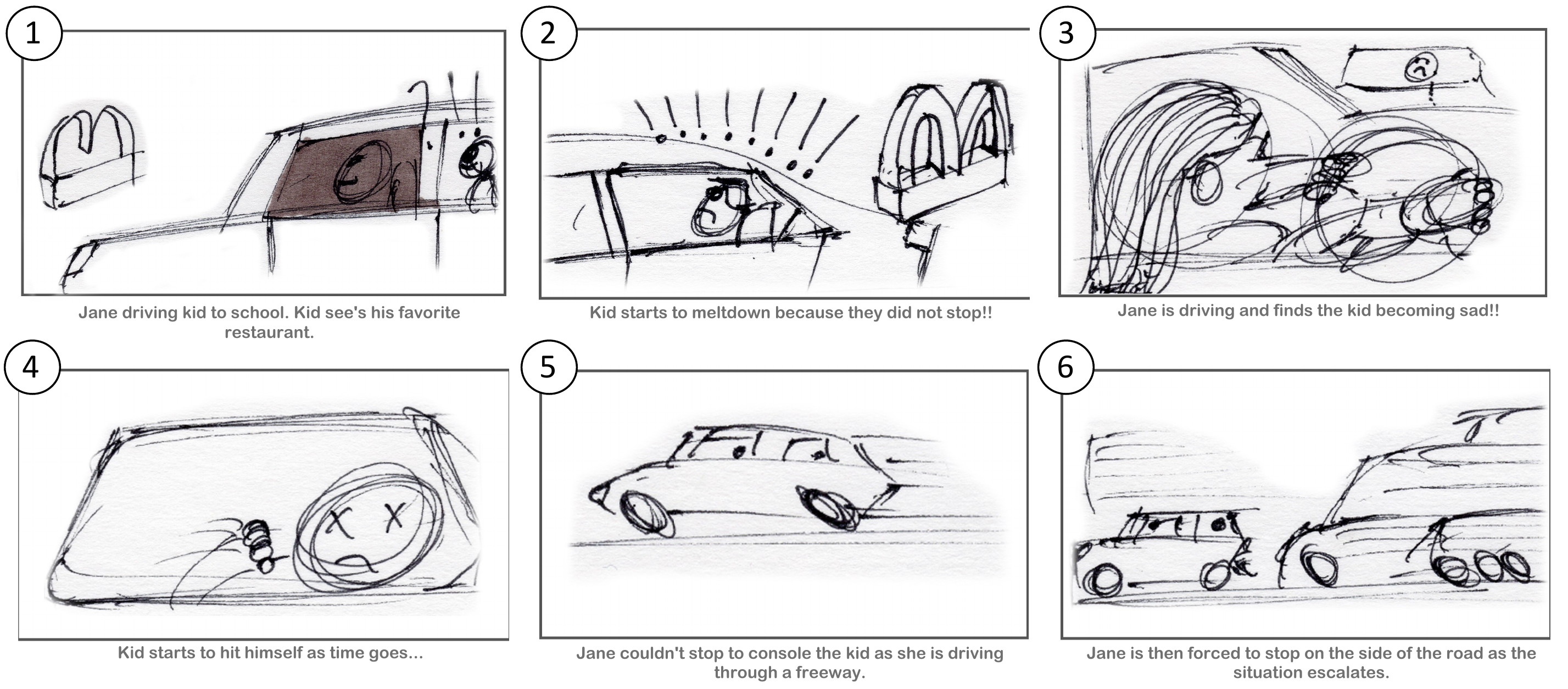}
  \caption{Storyboard that describes the first scenario.}
  \label{fig:scenario}
    \vspace{-10pt}
\end{figure*}

Based on these requirements, two personas and two scenarios using storyboards were created in order to understand users' goals, motivations, needs, pain points and more in the context of how they would use a possible design solution in order to generate ideas in the following stage. The first persona was created for a teenage with ASD and the second one was for a caregiver who shares the same ride with someone with ASD. The first scenario portrayed a caregiver who passed by an area that triggered the passenger with ASD with negative emotions and unsafe behaviors by forcing the caregiver to pull off the road to help. The second scenario portrayed how an ADS-DV transported the a passenger with ASD safely from place A to place B. As an example, we showed the first scenario in Figure \ref{fig:scenario}. These design activities allowed us to pin point the most important user needs and goals of the target users and how an ADS-DV could potentially be designed to ease both parties out of situations, such as providing more support to the rider, trustworthy safety measures, ways to shift those with ASD to focus on something more pleasant during the rides. 

\subsection{Ideate}
By examining the results from the previous two phases, we needed to generate a large number of ideas that could potentially satisfy the needs of the two target user groups by designing an ADS-DV. We went back to the affinity diagram and wrote down as many ideas as possible on sticky notes using Google Jamboard. We then conducted a brainstorming session among the 6 team members to identify the most promising ideas.  Based on the two major personas, we grouped the ideas into two categories, one for designing the interior of the AV and one for developing the mobile application to improve the environment of the vehicles and communicate between the passenger and his/her caregivers. The mobile application was mainly motivated by the low level of trust that parents or caregivers of those with ASD would let them drive in the AV alone (see Table 1). By designing the mobile application, it not only was used to help book a ride with the AV and provide preferred features (e.g., to keep them entertained, communicate with their families at any time), but also was used to monitor the drive in real time. The interior of the AV was motivated to keep them safe and comfortable with specially designed constraints and interfaces.

Likewise, we split the team into two subteams to help further develop the ideas into concepts. Once splitting up the team, we brainstormed again in designated groups to help fully develop our concepts. While each group talked among themselves, rough sketches were developed and then brought forward to the whole team for further discussions by checking on the personas and scenarios. We finally came up with the top idea which included an application for the caregivers to use as well as those with ASD to help create a soothing environment. We also proposed an interior design of the AV to provide a safe, stress-free environment. It is assumed that the targeted audience are expected to have little to no driving experience.

\subsection{Prototype and Test}
\begin{figure*} [bt!]
	\centering
	\subfloat[\label{fig:app1}]{\includegraphics[width=.525\linewidth]{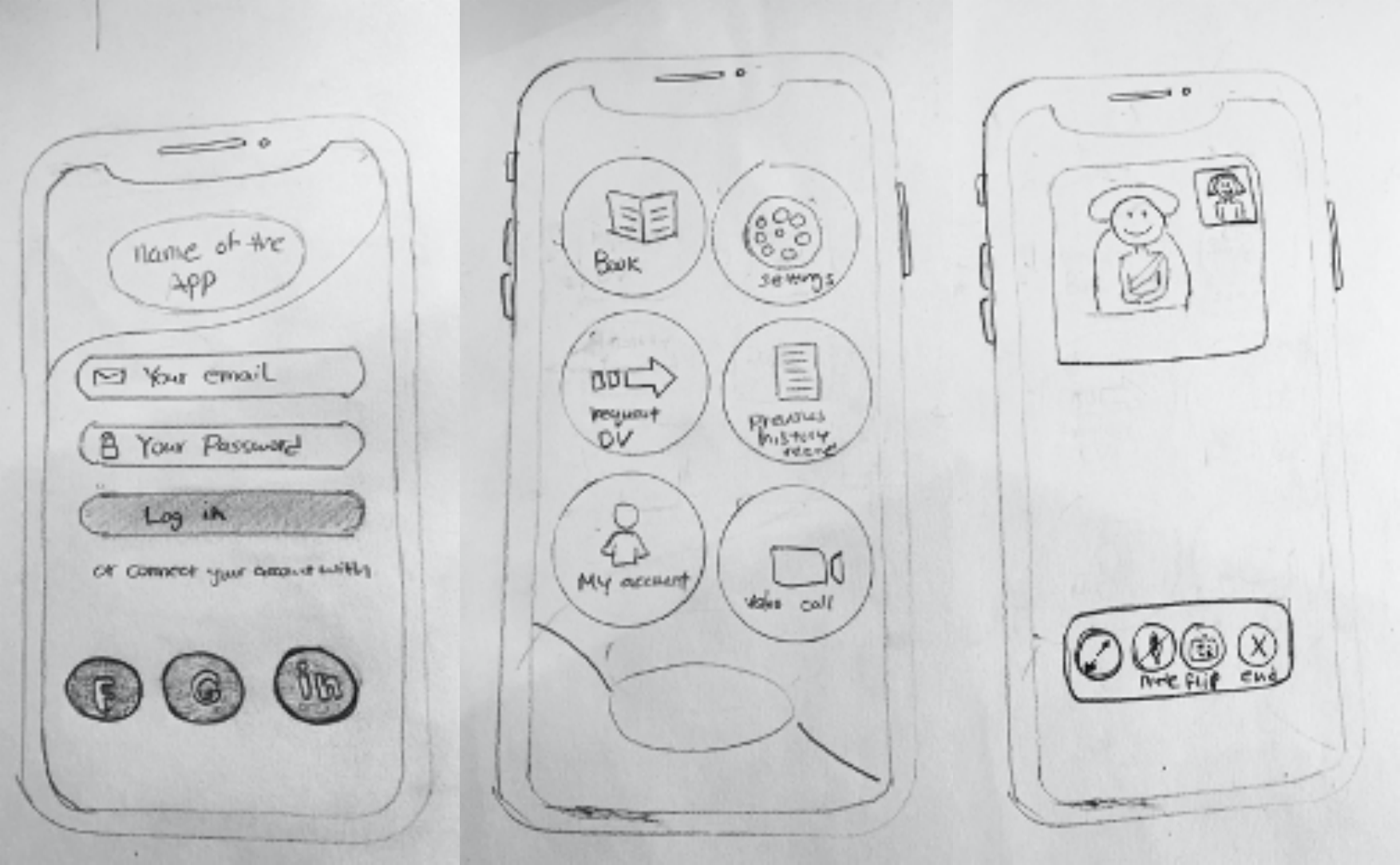}}
	\hspace{0pt}
	\subfloat[\label{fig:interior1}]{\includegraphics[width=.44\linewidth]{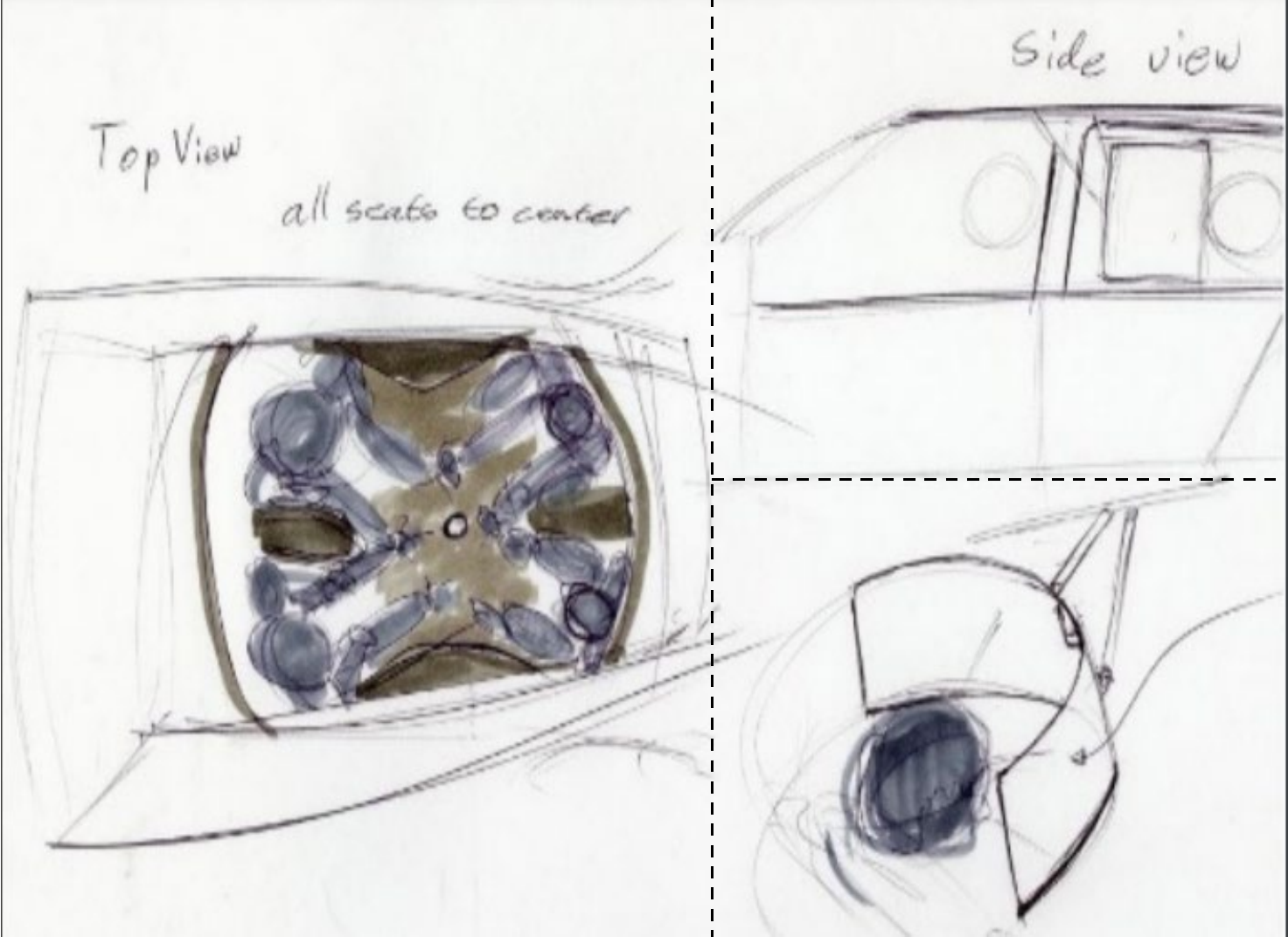}}
    \caption{Low-fidelity prototypes: (a) Sketches of the interfaces of mobile applications; (b) sketches of the interior of the ADS-DV.}\label{fig:lowfidelity}
\end{figure*}

In order to test the ideas generated in the previous stage, we started with low-fidelity prototypes by just sketching them on papers. Figure \ref{fig:lowfidelity} shows the prototypes of the mobile application interfaces and interior of the ADS-DV. The mobile application was also designed to empower people with ASD to get a ride on demand for appointments, monitor those with ASD in the drive to improve trust in AV, and to customize the vehicle with a desirable environment while the interior was to prevent the passengers from social distractors that could potentially elicit undesirable reactions. These low fidelity prototypes were tested with three participants in the empathy stage to get feedback. Two major concerns from the feedback were the interface tended to be complicated for those with ASD (see Figure \ref{fig:app1}) and the divider inside the vehicle might elicit negative emotions (see Figure \ref{fig:interior1}). Based on such feedback, we further refined the prototypes by simplifying the mobile applications and removing the divider. The participants were satisfied with these improvements, but wanted to see the interior with a high fidelity. 

\begin{figure*} [bt!]
	\centering
	\subfloat[\label{fig:app2}]{\includegraphics[width=.45\linewidth]{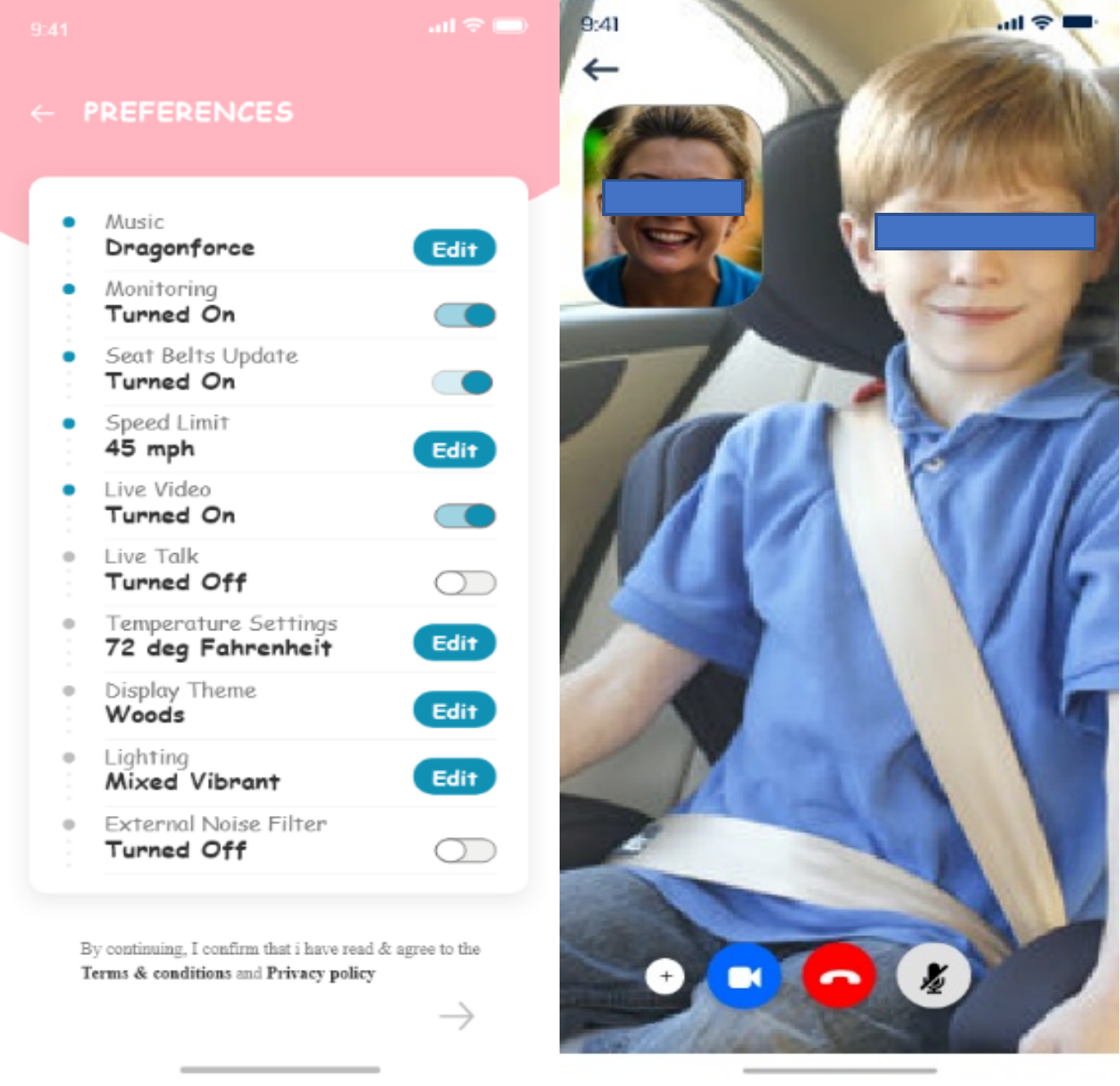}}
	\hspace{0pt}
	\subfloat[\label{fig:interior2}]{\includegraphics[width=.53\linewidth]{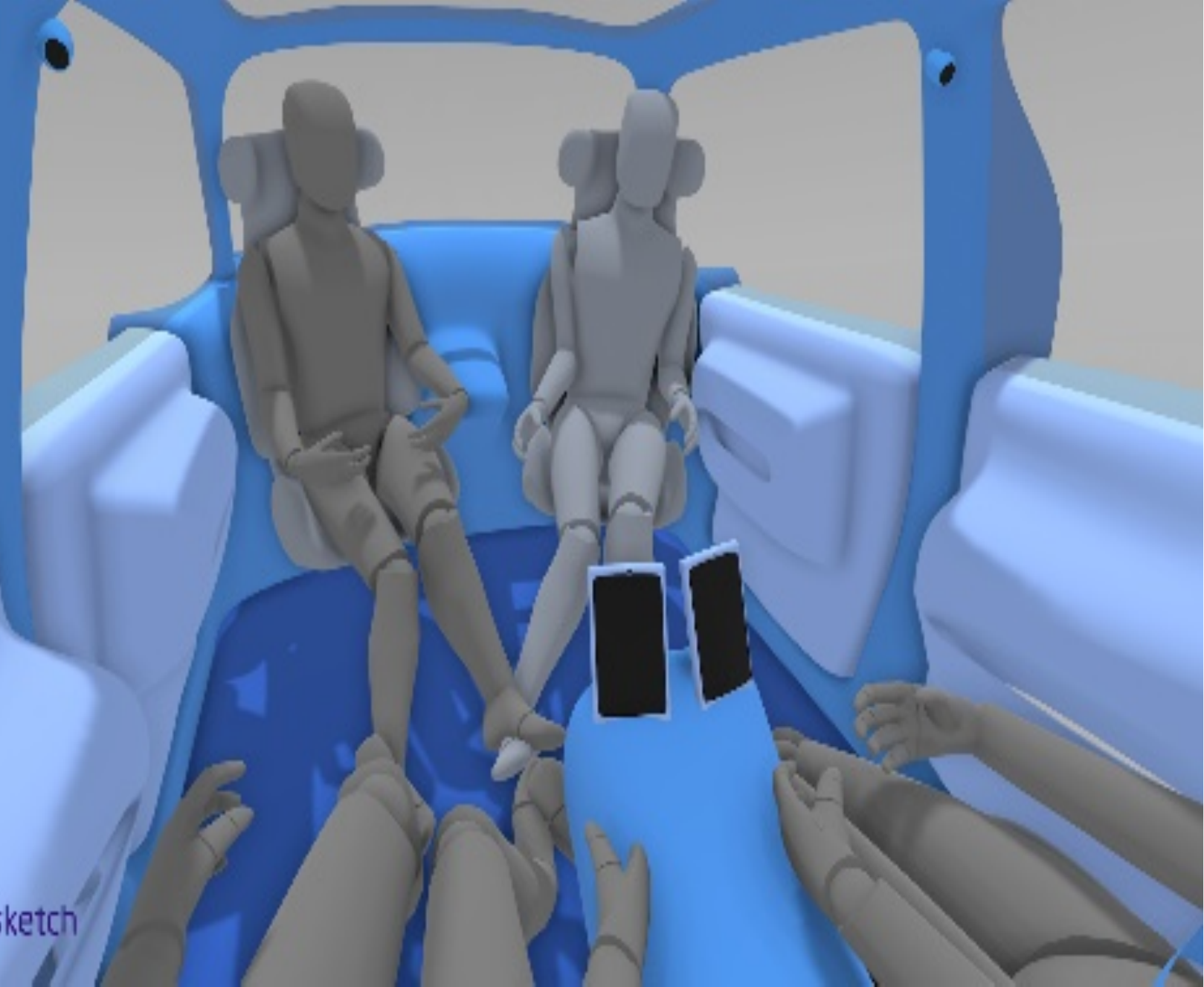}}
    \caption{High-fidelity prototypes: (a) Interfaces of mobile applications focusing on customizing preferences and video communication (see renders at https://tinyurl.com/driveotsi); (b) Virtual rendering of the interior of the ADS-DV.}\label{fig:highfidelity}
    \vspace{-10pt}
\end{figure*}

Then, we used Adobe XD (San Jose, California, U.S. www.adobe.com/products/xd.html) to generate a high-fidelity interface for the mobile application to focus on the customization function and video communication/monitoring. The customization function was deemed to be important due to the strong individual preferences of those with ASD and the video communication and monitoring function was also important because trust in fully automated driving could still be an issue \cite{zhou2020takeover,Ayoub2021Modeling}, let alone for people with ASD to travel without caregivers. In order to create a high-fidelity vehicle interior, we used a virtual reality modeling application called Gravity Sketch (London, U.K., https://www.gravitysketch.com/), which allowed us to render the interior with 3D mannequins and to view the interior from a first-person perspective of the passengers. Figure \ref{fig:highfidelity} shows the high-fidelity prototypes. 

\begin{figure*} [bt!]
	\centering
	\subfloat[\label{fig:interiorWindow1}]{\includegraphics[width=.45\linewidth]{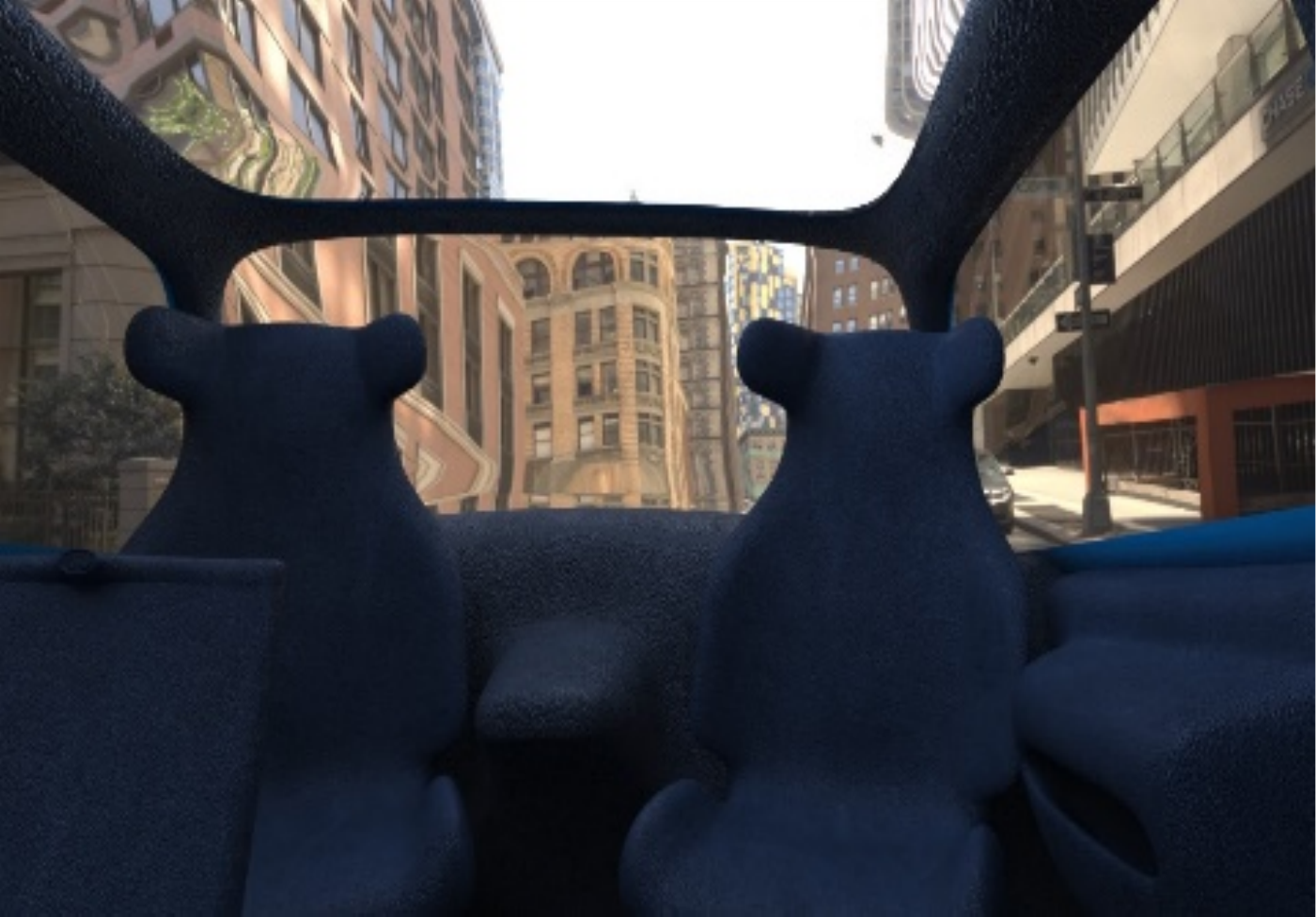}}
	\hspace{0pt}
	\subfloat[\label{fig:interiorWindow2}]{\includegraphics[width=.45\linewidth]{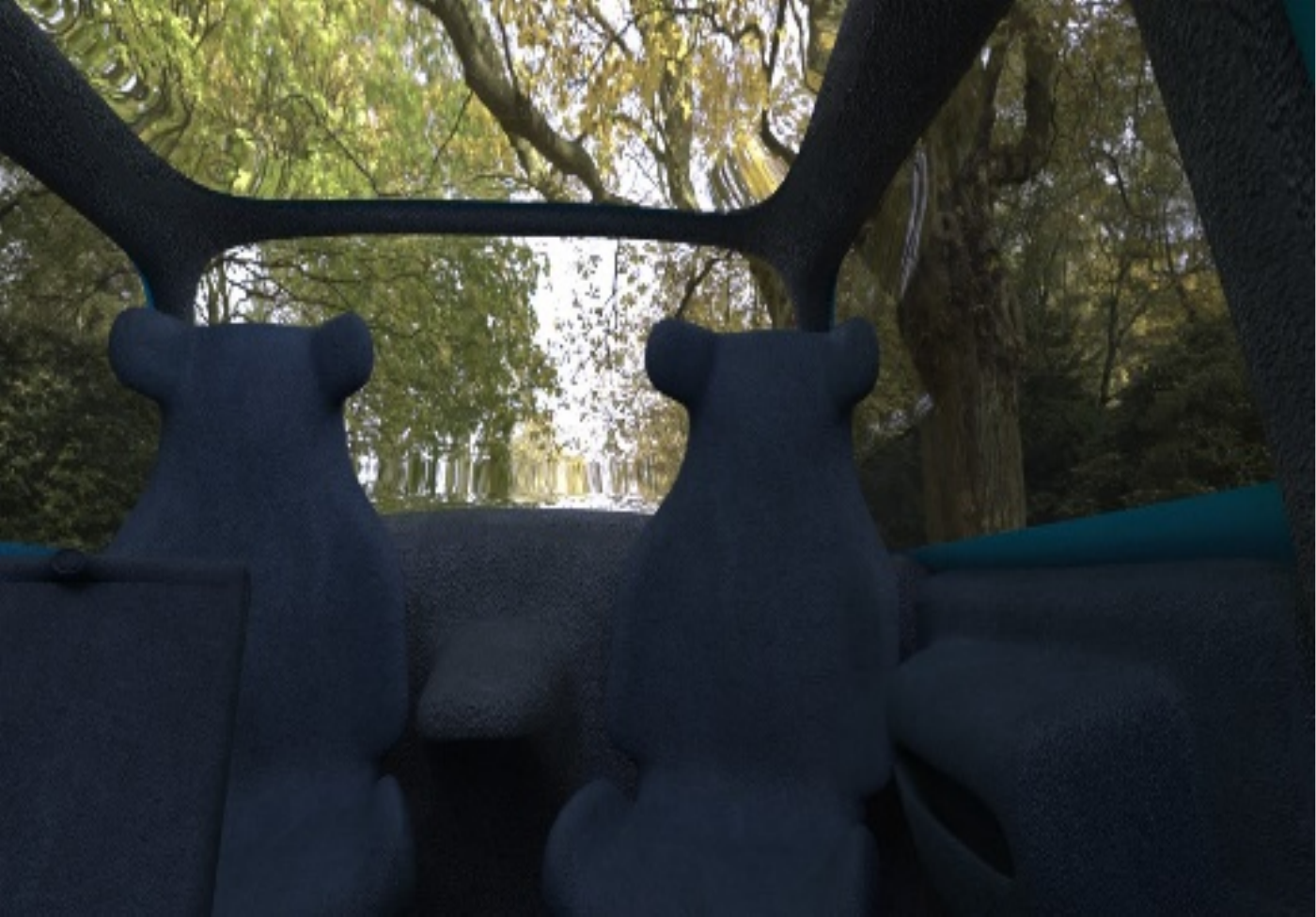}}
    \caption{Virtual rendering on vehicle windows (a) City view without possible social distractors; (b) Country view without possible social distractors.}\label{fig:interiorWindow}
    \vspace{-10pt}
\end{figure*}

These high-fidelity prototypes were further tested. We invited the three participants from the empathy stage to be involved in a co-creation session so that the important features were included and possible issues were resolved in this process. One key feedback was that people with ASD might not like to view the social distractors outside the vehicle, especially in urban and suburban areas. Moreover, many people with ASD often have strong preferences on what they like to see during the ride.  In order to address these issues, we further rendered the virtual stimuli through the windows of the vehicle based on the preferences of individual passengers. Two examples are shown in Figure \ref{fig:interiorWindow}. These visual stimuli can be set by the mobile application.

\section{Discussions and Conclusions}
Our research results showed that transportation for passengers with ASD faced tremendous difficulties when commuting from one place to another. These struggles ranged from sensory behavior issues, lack of understanding of traffic signals, and inability to focus on a task for a prolonged period of time. This motivated us to design an ADS-DV that would address the mobility needs and goals of the passengers with ASD and their caregivers. Taking into consideration the aforementioned results, we designed a high-fidelity prototype AV which was then used to mimic the “safe interior” for passengers with ASD. Conclusively, it was deemed that there was a dire need for vehicles to be designed to tailor the needs of passengers with ASD. Based on the feedback, we found

\begin{enumerate}
\item{All the participants agreed that riding in an AV with SAE Levels 4-5 would significantly improve the driving experience for passengers with ASD.}
\item{Combination of the mobile application and the new interior design would allow the caregiver to improve their trust in ADS-DV and to provide a comfortable ride for people with ASD alone.}
\item{The configuration of the interior design was helpful to overcome difficulties with current vehicles.}
\item{Design considerations in seat belts and headsets were well enough to prevent self-injury.}
\item{Customization options was easy to use.}
\item{Additional tablet devices in the armrest was more preferable and easy to communicate between caregivers and passengers and to monitor the behaviors of the passengers.}
\item{The virtual rendering on the windows provided further customization to help create a soothing and pleasant ride.}
\end{enumerate}

This study also has limitations. Firstly, the sample size involved in the study was rather small, where out of the 31 participants, only 2 were with ASD. Moreover, we could not involve people with ASD in the testing phase. Due to these limitations it was difficult to fully understand how the prototype satisfied the target user needs. Secondly, the entire study was done virtually due to the COVID-19 pandemic. We did the testing via Zoom which reduced the in-person interaction and in-depth and iterative human-centered design procedures, as well as did not allow participants to experience the effect of virtual environment and evaluate usage experience of the mobile application. However, we plan to explore and find answers to these issues in future research which should be conducted with a possible physical prototype, and will invite more participants to have interactive experience with the prototype.
%Despite the positive feedback from the participants, it is also important to recognize the limitations of this study.the entire study was based on the assumption of using SAE Levels 4-5 AVs, which are not in production yet and also the participants had no experience in such scenarios which made it difficult for them to interactively experience the prototypes. Secondly, the entire study was done virtually due to the COVID-19 pandemic which reduced the in-person interaction and in-depth and iterative human-centered design procedures that could have been carried out in the normal times. 
%Thirdly, since the targeted audience of the study were a specific category of users, the sample size involved in the study was rather small, where out of the 31 participants people with ASD were only 2. Further research should be conducted with a possible physical prototype to invite more participants (both people with ASD and caregivers of them) to have interactive experience with the prototype.

\bibliographystyle{ACM-Reference-Format}
\bibliography{main}

\end{document}